\documentclass[craaa]{baaa}

\usepackage[pdftex]{hyperref}
\usepackage{helvet}
\usepackage{afterpage}
\usepackage[pdftex]{hyperref}
\usepackage{subfigure}
\usepackage{natbib}
\usepackage{helvet,soul}
\usepackage[font=small]{caption}
\newcommand{\chandra}{{\it Chandra}}

\newcommand{\tess}{\textit{TESS}}
\newcommand{\gaia}{{\it Gaia}}

\newcommand{\nova}{V1674~Her}

\contriblanguage{1}

\contribtype{1}

\thematicarea{13/17}

\title{
The orbital period of the nova V1674~Her as observed with \tess}

\author{
G. J. M. Luna\inst{1}, I. J. Lima\inst{2,3}, M. Orio\inst{4,5}
}

\institute{
CONICET-Universidad Nacional de Hurlingham, Av. Gdor. Vergara 2222, Villa Tesei, Buenos Aires, Argentina \and
CONICET-Universidad de Buenos Aires, Instituto de Astronomía y Física del Espacio (IAFE), Av. Inte. Güiraldes 2620, Buenos Aires, C1428ZAA, Argentina \and
Universidad Nacional de San Juan, Facultad de Ciencias Exactas, Físicas y Naturales, Av. Ignacio de la Roza 590 (O), Complejo Universitario ``Islas Malvinas", Rivadavia, J5402DCS, San Juan, Argentina \and
Department of Astronomy, University of Wisconsin, 475 N. Charter Str., Madison, WI, USA \and
INAF-Padova, Vicolo Osservatorio 5, I-35122 Padova, Italy
}

\abstract{
Nova Her 2021 was observed with \tess\ 12.62 days after its most recent outburst in June 12.537 2021. This cataclysmic variable belongs to the intermediate polar class, with an spin period of $\sim$501 s and orbital period of 0.1529 days. During \tess\ observations of Sector 40, the orbital period of 0.1529(1) days is detected significantly 17 days after the onset of the outburst. A modulation, of unknown origin, with a period of $\sim$0.537 days is present in the data from day 13 until day 17.
}

\keywords{
binaries: close, (stars:) novae, cataclysmic variables, stars: individual (V1674 Her)
}

\contact{juan.luna@unahur.edu.ar}

\begin{document}
\maketitle
\section{Introduction}

V1674~Her (Nova Her 2021) was reported in outburst on 2021 June 12.537 UT at 8.4 magnitudes by Seiji Ueda (Kushiro, Hokkaido, Japan) reaching naked-eye magnitudes at its peak (\citealt{2021ATel14704....1M,2022ATel15796....1M} and other references in \citealt{2021ApJ...922L..42D}). Early X-ray observations obtained with \chandra\ show an strong modulation with a period of 503.9 s \citep{2021ATel14776....1M}. A period of 501.42~s was later reported by \cite{2021ATel14720....1M} found in ZTF data obtained during quiescence. These were the firsts hints of the presence of a magnetic accreting white dwarf in the system. The period measurement was refined and confirmed in subsequent X-ray monitoring \citep{2021ATel14798....1P, 2021ApJ...922L..42D, Orio2022} and was clearly detected during all the supersoft X-ray phase \citep{Orio2022}. 

Photometric data obtained by \cite{2021ATel14835....1S} and \cite{2021ATel14856....1P} allowed the detection of a 0.15302(2)~days period in optical wavelengths. This has since then been identified as the orbital period. Then, V1674~Her has the physical characteristics to be identified as a magnetic cataclysmic variable of the intermediate polar type. The orbital period was later refined to a value of 0.152921(3) days by \citet{2022ApJ...940L..56P} by analyzing fast photometry optical data. A 0.153 days period likely present in X-ray data has been reported by \cite{2022MNRAS.517L..97L}.

\section{Observations} 

\begin{figure*}
\includegraphics[scale=0.8]{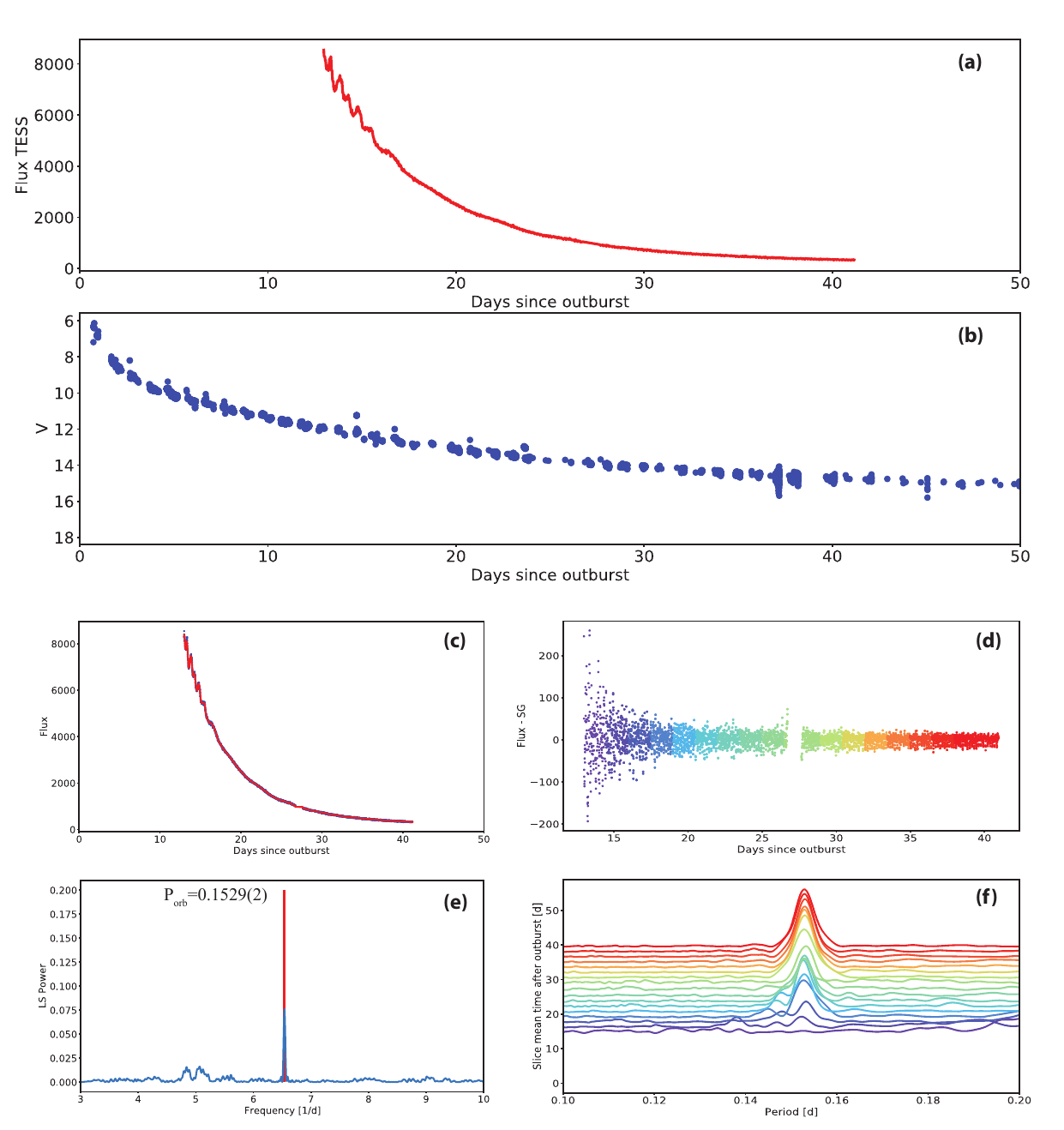} \caption{{\em Panel (a):} \tess\ light curve of \nova\ observed during Sector 40, 13 days after outburst. {\em Panel (b):} AAVSO, V-magnitud light curve of \nova\ after outburst on June 12 2021. {\em Panel (c):} \tess\ light curve with a SG filter (red line) to remove the long term trend.  {\em Panel (d):} \tess\ light curve after applying the SG filter. {\em Panel (e):} Lomb-Scargle power spectrum of the \tess~-SG light curve. The red vertical line marks the frequency of the orbital period at P$_{orb}$=0.1529(1) days.  {\em Panel (e):} Lomb-Scargle power spectra of each 3-day portion of light curve. Colors correspond to the colors in Panel (d). The 3-day portions of the light curve were extracted from a moving window with 50\% overlapping.  The orbital period is significantly detected after day $\sim$ 17.}
         \label{Fig:lc}
\end{figure*}

\nova\ was observed with \tess\ during Sector 40, which started on June 25 2021, 12.62 days after outburst. The total observing time was 676~h. Figure~\ref{Fig:lc}a shows the \tess\ light curve in the context of the post-outburst light curve from AAVSO (Fig.~\ref{Fig:lc}b). We extracted the TESS Full Frame Images (FFI) with a cadence of 10~min using {\tt TESSCut} tool \citep{Brasseur_2019} from the Python package {\tt lightkurve} \citep{Lightkurve_2018}. 
The target was identified by its {\tt SIMBAD} coordinates and \gaia\ catalog. We tested different aperture masks thresholds and background masks in order to produce the 
light curve.
The images were selected using hard QUALITY flag discarding unwanted events. We removed the long-term trend, due to the nova fading, by substracting a Savitzky-Golay filter \citep{Savitzky_1964} (see Figure \ref{Fig:lc}c). 

We searched for periods using the Lomb-Scargle algoritm in the light curve with the SG filter applied. We also searched for changes in the orbital period by extracting the power spectrum in 3-days-long sliding windows that overlap 50\% (Figure \ref{Fig:lc}d). 

\section{Results}

\begin{figure}[ht!]
\includegraphics[scale=0.5]{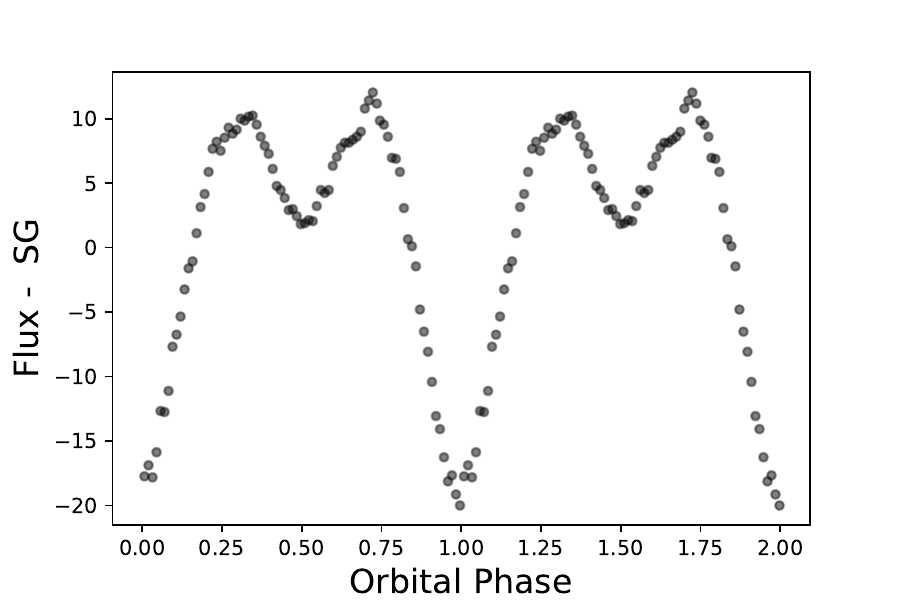} 
\caption{\tess\ light curve folded at the orbital period P$_{orb}$=0.1529 days. The light curve was binned at 80 bins/period.}
\label{Fig2}
\end{figure}

The \tess\ light curve of \nova\ from day 17 after outburst is modulated at the orbital period of P$_{orb}$=0.1529(1) days, consistent with previous findings. The light curve folded at the orbital period and taking T$_{0}$ from \citet{2022ApJ...940L..56P}, shows a double-peak profile (see Fig. \ref{Fig2}), with secondary minima distant at 
$\Delta\phi$=0.5 from the main minimum. 

Although the observations started 13 days after the outburst, remarkably, the orbital period was not detected until $\sim$17 days after the outburst, also in agreement with \citet{2022ApJ...940L..56P}. 
We found that within the error of 0.001 days, we could not detect any changes in the orbital period during the 28.2 days covered by \tess. \citet{2022ApJ...940L..56P} reports a systematic drift toward longer periods in the photometric data taken in 2022 with respect to those taken in 2021. Our observations do not cover the times when the source has returned to quiescence, when the changes in the orbital period seem to manifests.

\begin{figure}[ht!]
\includegraphics[scale=0.5]{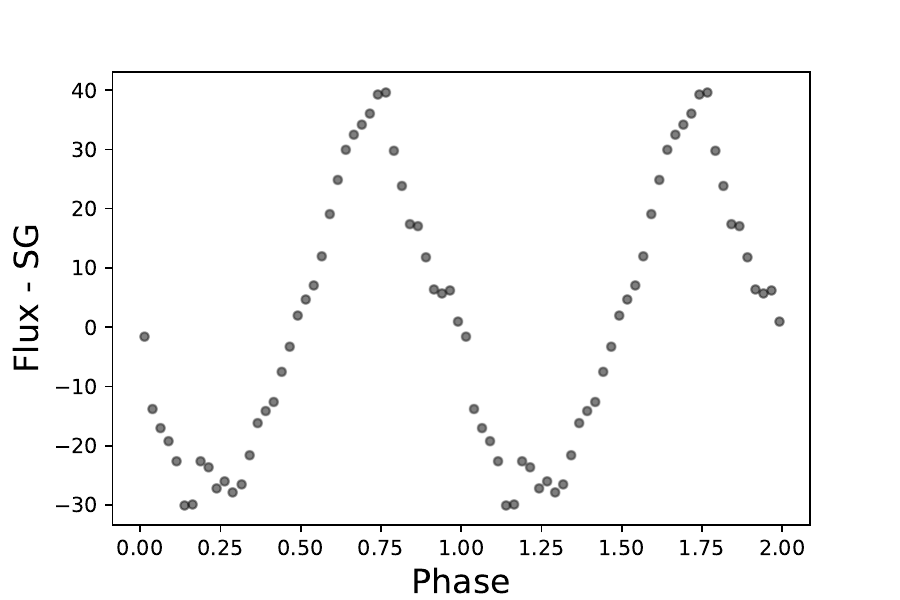}
\caption{\tess\ light curve folded at period P$_{orb}$=0.537 days with arbitrary ephemeris. The light curve was binned at 40 bins/period.}
\label{Fig3}
\end{figure}

The variability on time scales longer than the orbital period reported by \citep{2023MNRAS.521.5453S} is also observed in the \tess\ light curve from day 13th until day 17th. A search for periods in this segment of the light curve yields a period of $\sim$0.537 days, of unknown origin (see Fig. \ref{Fig3}). This period is unlikely to be due to superhumps, which are modulations with periods slightly below or above the orbital period and likely due to a warped/precessing accretion disk. The 0.537 days periods is much longer than the orbital period.

\section{Conclusion}

We analyzed the \tess\, observations of Nova Her 2021, that entered into Sector 40 12.62 days after being reported to be in outburst. This nova belongs to the intermediate polar class of cataclysmic variables and as such, it displays clear modulations with the orbital and WD spin periods. In agreement with other authors, we found that the light curve shows two minima, separated by half a cycle. The secondary minimum is most noticeable once the light curve is decaying more slowly and the systems is reaching quiescence conditions. This secondary minimum could be due to the eclipsed, enhanced emission due to irradiation of the secondary by the still-burning white dwarf \citep{2022ApJ...924...27P}.

\begin{acknowledgement}

GJML and IJL acknowledge support from grant ANPCYT-PICT 0901/2017. GJML is a member of the CIC-CONICET (Argentina). This research made use of Lightkurve, a Python package for Kepler and TESS data analysis (Lightkurve Collaboration, 2018). We also acknowledge the variable star observations from the {\sc AAVSO} International Database contributed by observers worldwide and used in this research.

\end{acknowledgement}

\bibliographystyle{baaa}
\small
\bibliography{sample631}

\end{document}